\documentclass[aps,prc,twocolumn,floatfix,nofootinbib,preprintnumbers,superscriptaddress,longbibliography]{revtex4-1}

\usepackage{epsfig}
\usepackage{graphicx}
\usepackage{stmaryrd}
\usepackage{amssymb,tabularx,dcolumn}
\usepackage{supertabular,ltxtable}
\usepackage{longtable}
\usepackage{amsmath}
\usepackage{amstext} 
\usepackage{float}
\usepackage{color}
\usepackage{bm}
\usepackage{CJKutf8}
\usepackage{hyperref}
\usepackage{mathtools}
\usepackage{multirow}
\usepackage{makecell}
\usepackage[utf8]{inputenc}
\usepackage{tabu}
\usepackage{braket}
\usepackage{ulem}

\makeatletter
\def\@bibdataout@aps{%
\immediate\write\@bibdataout{%
@CONTROL{%
apsrev41Control%
\longbibliography@sw{%
    ,author="08",editor="1",pages="1",title="0",year="1"%
    }{%
    ,author="08",editor="1",pages="1",title="",year="1"%
    }%
  }%
}%
\if@filesw \immediate \write \@auxout {\string \citation {apsrev41Control}}\fi 
}
\makeatother

\renewcommand{\vec}[1]{\mbox{\boldmath $#1$}}

\renewcommand{\vec}[1]{\mbox{\boldmath $#1$}}

\usepackage{hyperref}

\definecolor{star}{RGB}{31,119,180}

\begin{document}
\begin{CJK*}{UTF8}{gbsn}
\date{Received 22 July 2021; revised 31 August 2021; accepted 16 November 2021; published 3 December 2021}

\title{Spectroscopic factors in dripline nuclei}

\author{J. Wylie}
\affiliation{FRIB/NSCL Laboratory, Michigan State University, East Lansing, Michigan 48824, USA}

\affiliation{Department of Physics and Astronomy, Michigan State University, East Lansing, Michigan 48824, USA}

\author{J. Oko{\l}owicz}
\affiliation{Institute of Nuclear Physics, Polish Academy of Sciences, Radzikowskiego 152, PL-31342 Krak{\'o}w, Poland}

\author{W. Nazarewicz}
\affiliation{FRIB/NSCL Laboratory, Michigan State University, East Lansing, Michigan 48824, USA}
\affiliation{Department of Physics and Astronomy, Michigan State University, East Lansing, Michigan 48824, USA}

\author{M. P{\l}oszajczak}
\affiliation{Grand Acc\'el\'erateur National d'Ions Lourds (GANIL), CEA/DSM - CNRS/IN2P3, BP 55027, F-14076 Caen Cedex, France}

\author{S.M. Wang (王思敏)}
\affiliation{FRIB/NSCL Laboratory, Michigan State University, East Lansing, Michigan 48824, USA}
\affiliation{Institute of Modern Physics, Fudan University, Shanghai 200433, China}

\author{X. Mao (毛兴泽)}
\affiliation{FRIB/NSCL Laboratory, Michigan State University, East Lansing, Michigan 48824, USA}
\affiliation{Department of Physics and Astronomy, Michigan State University, East Lansing, Michigan 48824, USA}

\author{N. Michel}
\affiliation{Institute of Modern Physics, Chinese Academy of Sciences, Lanzhou 730000, China}
\affiliation{School of Nuclear Science and Technology, University of Chinese Academy of Sciences, Beijing 100049, China}


\begin{abstract}
 Single-nucleon knockout reaction studies 
 of the proton-dripline nuclei $^9$C and  $^{13}$O suggest an appreciable suppression of spectroscopic factors.  In this work, we calculate the one-neutron and one-proton spectroscopic factors for the mirror pair $^9$C-$^9$Li and $^{13}$O using two variants of the continuum shell model: the complex-energy Gamow Shell Model  and the real-energy Shell Model Embedded in the Continuum. Our results indicate that the continuum effects strongly suppress the spectroscopic factors of well-bound orbits in the dripline systems, but have less impact on the spectroscopic factors of weakly-bound states.
\end{abstract}

\maketitle
\end{CJK*}


{\it Introduction}. Spectroscopic factors (SFs) extracted from $(e,e'p)$ experiments yield the $\sim$35\% quenching with respect to the shell model values \cite{Lapikas1993}. Several mechanisms have been put forward to explain this reduction. These include the effects of short-range  and  long-range correlations  \cite{Dickhoff2010,Paschalis2020,Aumann2021}.
In a series of papers \cite{Gade2004, Gade2008,Tostevin2021}, it was found that the ratio $R_s =\sigma_{\rm expt}/\sigma_{\rm theor}$ of the experimental and theoretical  inclusive one-nucleon removal cross section for a large number of projectiles shows a strong dependence on the $\Delta S =S _p - S_n$ asymmetry of the neutron and proton separation energies. 
Numerous papers have discussed the isospin dependence of SFs \cite{Barbieri2009,Flavigny2012,Grinyer2012,Kay2013,Timofeyuk2013,Tostevin2014,Lee2016,WangST2018}, or lack of it \cite{Atar2018,Gomez2018,Phuc2019}, see the recent review \cite{Aumann2021} for a comprehensive discussion and additional references. 

However, in spite of significant efforts, the problem remains. Experimental studies of $(p,2p)$ reactions on oxygen and carbon isotopes did not find a significant dependence of SFs on proton-neutron asymmetry \cite{Atar2018, Holl2019}. A similar conclusion has been made in theoretical studies of neon isotopes and  mirror nuclei $^{24}$Si, $^{24}$Ne and $^{28}$S, $^{28}$Mg \cite{Okolowicz2016}. On the other hand,  coupled-cluster studies \cite{Jensen2011} of the neutron-rich oxygen isotopes have shown a significant quenching of SFs.
 In light nuclei ($A=7-10$)  it has been shown \cite{Grinyer2012} that the separation between nuclear reaction and structure, as given by the eikonal reaction model \cite{Gade2008} and the Variational Monte Carlo structure model \cite{Wuosmaa2008}, provides reasonable description of experimental one-nucleon knockout cross sections only when nucleons with large separation energies are removed. 
As concluded in \cite{Tostevin2021}, the physical origins of the presented systematic behavior of $R_s$ versus $\Delta S$ remain
unresolved.

Based on the  dispersive optical model 
analysis of  proton scattering data  \cite{charity2006, charity2007}, one might hypothesise that the essential ingredient behind the systematics of $R_s$  is the dispersion relation which connects real and imaginary parts of the scattering matrix. Following this line of reasoning, one can argue that the understanding of the $R_s(\Delta S)$ systematics should be related to the theoretical treatment of the nuclear openness, i.e.,  to a  treatment of the coupling between bound states, resonances, and the non-resonant (scattering) continuum.

Let us emphasize that while the experimental SFs are deduced from measured nucleon transfer cross sections by using reaction models. theoretical SFs are obtained from nuclear structure calculations by computing overlap integrals or spectroscopic amplitudes. Reaction and structure models used to describe  SFs are usually not consistent, i.e.,  different parts entering the theoretical cross-section calculations are usually based on different frameworks or assumptions. Only when all aspects of calculations  are used in a consistent framework, can a meaningful comparison to experiment be made. This has not yet been accomplished.

The objective of  this Letter is to investigate the role of the continuum coupling on SFs in weakly-bound and unbound nuclei.
To this end, by means of open-quantum-system configuration-interaction frameworks, we study one-proton and one-neutron removal SFs from
proton-rich  $^9$C,  $^{13}$O, and $^{13}$F, and neutron-rich $^9$Li. This choice has been driven by the recent experimental studies \cite{Charity2020,Charity2021}.

{\it Method}. To provide a comprehensive description of  continuum coupling effects, in this Letter, we adopt two different open-quantum-system frameworks: the Gamow Shell Model (GSM) \cite{Michel2002,Michel2009} and the Shell Model Embedded in the Continuum (SMEC) \cite{Bennaceur2000,Okolowicz2003},
which have been successfully used for studies of  weakly bound and unbound states in   dripline nuclei. 
Both frameworks describe the nucleus as a core surrounded by valence nucleons, but they treat the coupling to the unbound  continuum space differently. 
In the GSM, the continuum effects are automatically taken into account by utilizing the Berggren ensemble \cite{Berggren1968} that contains resonant (bound and decaying) and scattering states. In the SMEC, based on the Feshbach projection technique, the continuum space 
consists of one nucleon occupying scattering states.
By comparing the calculated SFs with the results obtained by the standard closed-quantum-system  (CQS) shell model,  the continuum effects on SFs can be  quantified.

The Hermitian GSM Hamiltonian can be written as a sum of
the kinetic energy of valence nucleons, the one-body core-valence interaction $\hat{U}_{\rm c}(i)$,  the two-body interaction $\hat{V}_{i,j}$, and the two-body recoil term:
\begin{eqnarray}
      {\hat H} &=& \sum_{ i } \left( \frac{ \vec{ p }_i^{ 2 } }{ 2 { \mu }_{ i } } + {\hat U}_{\rm c}(i) \right) + \sum_{ i < j } {\hat V }_{i,j}    \nonumber \\
     &+& \frac{ 1 }{ { M }_{\rm core} } \sum_{ i < j  } \vec{ { p } }_{ i } \cdot \vec{ { p } }_{ j }  ~ \ ,
      \label{H_COSM} 
    \end{eqnarray}
where $i, j=1,\dots N_v$ and  $N_{v}$ is the number of valence nucleons.
The recoil term, resulting from the change of Hamiltonian from the laboratory coordinates to the cluster-orbital shell model relative coordinates  \cite{Suzuki1988},
 is used to restore the translational invariance.
The treatment of this term follows the harmonic oscillator expansion procedure described in 
Ref.~\cite{Michel2010}. The GSM Hamiltonian is diagonalized in  the Berggren basis $|k_p\rangle$. In this way,  the continuum couplings between Slater determinants  involving bound and unbound nucleons are automatically taken into account.

In a multi-channel version of SMEC,  which is used here, the Hilbert space is divided into two orthogonal subspaces ${\cal Q}_0$ and ${\cal Q}_1$ containing 0 and 1 particle in the scattering continuum, respectively.
The  energy-dependent effective Hamiltonian of SMEC
\begin{equation}
{\cal H}(E)=H_{{\cal Q}_0{\cal Q}_0}+W_{{\cal Q}_0{\cal Q}_0}(E),
\label{eq21}
\end{equation}
can be decomposed into the CQS shell-model (SM) Hamiltonian $H_{{\cal Q}_0{\cal Q}_0}$ acting in ${\cal Q}_0$ and the  continuum coupling term:
\begin{equation}
W_{{\cal Q}_0{\cal Q}_0}(E)=H_{{\cal Q}_0{\cal Q}_1}G_{{\cal Q}_1}^{(+)}(E)H_{{\cal Q}_1{\cal Q}_0},
\label{eqop4}
\end{equation}
where $G_{{\cal Q}_1}^{(+)}$ is the one-nucleon Green's function and $H_{{\cal Q}_0{\cal Q}_1}$ and $H_{{\cal Q}_1{\cal Q}_0}$
represent the couplings between  subspaces ${\cal Q}_0$ and ${\cal Q}_1$.  The energy scale in (\ref{eq21}) is defined by the lowest one-nucleon emission threshold. 

There are two kinds of operators in $W_{{\cal Q}_0{\cal Q}_0}(E)$: 
  ${\cal O}_{il}^K = \langle a^{\dagger}_{i}{\tilde a}_{l} \rangle^K$, and 
  ${\cal R}_{kl (L) i}^{j_n}=\langle a^{\dagger}_{i} \langle{\tilde a}_{k} {\tilde a}_{l} \rangle^L \rangle^{j_n}$.
The matrix elements of ${\cal O}$ are calculated between the states in the ($A-1$)-particle system and, hence, couple different decay channels. The operators ${\cal R}$ act between different SM wave functions in the $A$ - and ($A-1$) - particle systems, i.e.,  are responsible for both the mixing of the SM wave functions in the nucleus $A$ and the coupling between decay channels. 

SMEC solutions in ${\cal Q}_0$  are found by solving the eigenproblem for the non-Hermitian effective Hamiltonian ${\cal H}_{{\cal Q}_0{\cal Q}_0}(E)$. The complex eigenvalues of ${\cal H}_{{\cal Q}_0{\cal Q}_0}(E)$ at energies $E_{\alpha}(E) = E$, determine the energies and widths of resonance states. In a bound system ($E<0$) the eigenvalues of ${\cal H}_{{\cal Q}_0{\cal Q}_0}(E)$ are real. In the continuum, $E_{\alpha}(E)$ corresponds to the poles of the scattering matrix.
Eigenstates $|\Psi_{A,\alpha}\rangle$ of ${\cal H}_{{\cal Q}_0{\cal Q}_0}(E)$ are linear combinations of SM eigenstates 
$|\Phi_{A,i}\rangle$ generated by the orthogonal transformation matrix $b_{A,\alpha i}(E)$.

 The center-of-mass in SM wave functions of SMEC is handled in the same way as in the standard SM, see \cite{Okolowicz2003}. The coupling to the continuum is calculated in the relative coordinates of the coupled-channel framework so no additional spuriosity is generated  beyond the one which may appear in the SM wave functions \cite{Okolowicz2003}.

{\it Spectroscopic factors}.--- While not observables in the strictest sense \cite{Furnstahl2002,Duguet2015,Gomez2021,Tropiano2021}, SFs are useful as  they capture information on configuration mixing in the many-body wave function. In GSM, SFs are defined in terms of spectroscopic amplitudes \cite{Michel2007,Michel2007a}
${\cal A}_{\ell j}(k_p) = \langle \Psi_{A} || a^+_{\ell j}(k_p) || \Psi_{A-1}\rangle/\sqrt{2 J_A + 1} $ :
\begin{equation} {\label{eq:spectroscopic_factor}}
{\cal S}^2_{\ell j} = \int\hspace{-1.4em}\sum {{\cal A}^2_{\ell j}(k_p)},
\end{equation}
where $\Psi_{A}$ is the wave function of the 
mass-$A$ system, $J_A$ is its total angular momentum, and $a^+_{\ell j}(k_p)$ is a nucleon  creation operator associated with the Berggren basis state $|k_p\rangle$. It is to be noted that  Eq.\,(\ref{eq:spectroscopic_factor}) involves the summation over discrete resonant states and integration along the contour of scattering states of the Berggren ensemble. In this way, ${\cal S}^2_{\ell j}$ is independent on the choice of the single-particle basis \cite{Michel2007}. In the GSM framework,  the complex conjugation arising in the dual space affects only the angular part and leaves the radial part unchanged, and this affects the definition of the scalar product in (\ref{eq:spectroscopic_factor}).

It is worth noting that  spectroscopic factors
(\ref{eq:spectroscopic_factor}) can be straightforwardly related to one-nucleon radial overlap integrals \cite{Michel2007a}. In the context of this paper, it is worth noting that 
the  asymptotic behavior of the one-nucleon overlap integral  can be associated with the complex generalized one-nucleon separation energy
$\tilde{S}_{1n}(A)\equiv S_{1n}(A)
-i/2 \left[\Gamma(A-1)-\Gamma(A)\right]$. Note that that the imaginary part of $\tilde{S}_{1n}$ naturally appears when
either parent or daughter nucleus is unbound.

Due to the coupling to one-nucleon decay channel(s), the SMEC eigenfunction $\Psi_{A,\alpha}$  is a linear combination of SM wave functions
$\Phi_{A,i}$: $\Psi_{A,\alpha} = \sum_i b_{A,\alpha i} \Phi_{A,i}$.
In the standard version of SMEC, dubbed SMEC1, the spectroscopic amplitude between SMEC state $\Psi_{\alpha,A}$ and the SM state $\Phi^i_{A-1}$
becomes:
${\cal A}^{i \alpha}_{\ell_j} = \sum_k{\cal A}^{ik}_{\ell_j} b_{A,\alpha k}$.  
By including  the continuum coupling in the $A-1$ nucleus, one obtains:
${\cal A}_{\ell_j}^{\beta\alpha} = \sum_{i,k} b_{A-1,\beta i} {\cal A}_{\ell_j}^{ik} b_{A,\alpha k}$. This version of calculations is referred to as SMEC2. The spectroscopic factor in SMEC is  defined as the sum of squared spectroscopic amplitudes associated with possible reaction channels. For instance, for the proton knockout 
$^{13}{\rm O}(3/2_{\rm g.s.}^-)\rightarrow {^{12}{\rm N}}(1^+_{\rm g.s.})$, both $p_{3/2}$ and $p_{1/2}$ partial waves contribute and their SFs are added.

Let us also mention that in both GSM and SMEC, the SFs can be related to the many-body asymptotic normalization coefficients \cite{Okolowicz2012}.

{\it Model space and parameters}. For the core, we took  the tightly bound $^4$He nucleus. 
The GSM Hamiltonian (\ref{H_COSM}) was defined as in Ref.~\cite{Jaganathen2017,Mao2020}. Namely, the 
core-nucleus  potential  $\hat{U}_{\rm c}(i)$ was taken in the  Woods-Saxon (WS) form (supplemented by a spin-orbit term and Coulomb potential), and a   Furutani-Horiuchi-Tamagaki (FHT) force \cite{Furutani1979}
was used 
to describe the two-body interaction $\hat{V}_{i,j}$.
The parameters for the potentials were taken from Ref.\,\cite{Mao2020} for the $ps$-shell model space. For comparison, $p$-shell SM calculations
in the harmonic oscillator basis (HO-SM)
were carried out with the same GSM Hamiltonian. The single-particle (s.p.) energies in the  HO-SM approximation were given by real parts of resonant states generating the GSM basis. These resonant states define the pole space.

In order to reveal the impact of higher-$\ell$ shells, we  extended the model space to the $psd$-shell. In this case, the core potential strength for protons  was readjusted until the ground-state (g.s.) energy of $^8$C was within 0.2\,MeV of the experimental value \cite{ensdf}. 
Once the model parameters  were determined for $^8$C, they were used for neutrons in the calculations for the mirror partners of $^{8,9}$C ($^8$He,$^9$Li).
The adjusted proton WS parameters were retained for $^8$B, $^9$C, and the neutron parameters were taken from Ref.\,\cite{Mao2020}. The same procedure was used to calculate the mirror pair ($^{8,9}$Li).
Due to the fact that $^{9}$C is particle-bound,  the $A=9$ nuclei   were calculated with only two particles allowed in the continuum space ($N_{\rm cont}$=2) as compared to $N_{\rm cont}$=4  for $A=8$.

The GSM pole space used in this work consisted of 
$0p_{3/2}$ and  $0p_{1/2}$ shell and also the $0d_{5/2}$ shell for $^8$B and $^8$Li. 
The complex-momentum contour defining the scattering  space
 was divided into 3 segments: $[0,k_{\rm peak}], [k_{\rm peak},k_{\rm mid}]$, and  $[k_{\rm mid},k_{\rm max}]$,  with the values 
$k_{\rm peak}=0.3$\,fm$^{-1}$, $k_{\rm mid}=0.4$\,fm$^{-1}$, and the cutoff momentum $k_{\rm max}=4$\,fm$^{-1}$.
Each segment was discretized with 5 Gaussian points.
The binding energies   and spectra of $A$ = 8,9 dripline nuclei obtained in our  GSM calculations are discussed in the supplemental material \cite{SM}.   The energy levels of mirror nuclei are reproduced fairly well, which suggest that the Coulomb energy displacement and the Thomas-Ehrman shift are under control in the GSM model.

In the SMEC calculations, the SM Hamiltonian  $H_{{\cal Q}_0{\cal Q}_0}$   was taken as the standard YSOX interaction \cite{Yuan2012} in the $4\hbar \omega$ ($psd$)-model space. The radial s.p. wave functions (in ${Q}_0$) and the scattering wave functions (in ${Q}_1$) are generated by the WS central potential supplemented by the spin-orbit and Coulomb terms
with the parameters of Ref.~\cite{Okolowicz2020}.
The continuum-coupling term $W_{{\cal Q}_0{\cal Q}_0}$ has been modeled by the Wigner-Bartlett contact interaction ~\cite{Okolowicz2020} described by two  physically relevant parameters: the overall continuum-coupling strength $V_0$ and the spin-exchange parameter $\alpha$ that can be used to study the isospin content of the  continuum coupling. Physically reasonable values of $|V_0|$ are in the interval 100-350\,MeV$\cdot$fm$^3$. As discussed in the  earlier papers
\cite{Luo2002, Michel2004, Charity2018a, Charity2018a} the value of the spin-exchange parameter
$\alpha = 2$ is appropriate for dripline nuclei, and we adopted this value in this Letter.


 To gain insights into the wave function fragmentation caused by the continuum coupling, 
 we shall study two situations: (i) knockout of well-bound, minority species, nucleons, i.e., neutrons (protons) from proton-(neutron-) rich nuclei and (ii) knockout of weakly-bound, majority species, nucleons,
 i.e., protons (neutrons) from proton-(neutron-)rich nuclei. For earlier studies of this problem, see coupled-cluster calculations in the Berggren basis  \cite{Jensen2011} and the  intranuclear-cascade model involving core excitations \cite{Louchart2011}.

{\it Knockout of well-bound nucleons}.
We first study the removal of the
$p_{3/2}$ neutron from the $J^\pi=3/2^-$ g.s. of 
$^9$C. Since the neutron separation energy in $^9$C is large,
the neutron is removed from a well-bound orbit.
Table~\ref{tab:SF-mirror-comparison} shows the SFs calculated  in GSM with different numbers of particles allowed to occupy  scattering states. The GSM results are  compared with the HO-SM calculations, in which the continuum effect is absent. In HO-SM, a SF of 0.86 is obtained, while this value becomes 0.67 in the GSM calculations when considering continuum coupling from the $ps$-shell. The inclusion of the $d$-shell leads to a further reduction of the SF down to ${\cal S}^2$ = 0.48 when four protons are allowed to occupy scattering states.
A significant reduction of the SF with respect to the HO-SM value is also predicted for the removal of the well-bound $p_{3/2}$ proton from the g.s. of $^9$Li.

\begin{table}[!htb]
\caption{Spectroscopic factors for the knockout of a $p_{3/2}$  nucleon from
the 3/2$^-$ g.s. of  $^9$C and $^9$Li to the g.s. of $^8$C,
$^8$He, $^8$B, and $^8$Li. The experimental neutron and proton separation energies \cite{ensdf} are shown (in MeV).  The GSM-$ps$ results were obtained  in the full $ps$  space while the GSM-$psd$ space additionally includes  scattering  $d$-waves.
The HO-SM result corresponds to the shell model calculation in the $0p$ space.
 $N_{\rm cont}$ is the number of particles allowed in non-resonant continuum of $A=8$ nuclei. The last row shows the contribution
 from the resonant $0p_{3/2}$ state.}
\begin{ruledtabular}
\begin{tabular}{ l  c  c  c  c  c  c }
\multicolumn{1}{c}{ Model } & $N_{\rm cont}$ & $^9$C$\rightarrow$$^8$C & $^9$Li$\rightarrow$$^8$He & $^9$C$\rightarrow$$^8$B & $^9$Li$\rightarrow$$^8$Li\\
  & & 14.22 & 13.94  & 1.30 & 4.06 \\
\hline \\[-4pt]
{HO-SM}   & 0 & 0.86 & 0.85 & 0.95 & 0.96 \\
{GSM-$ps$}  & 3 & 0.67 & 0.67 & 0.98 & 0.98 \\
{GSM-$psd$} & 3 & 0.60 & 0.67 & 0.89 & 0.88 \\
{GSM-$psd$} & 4 & 0.48 & 0.65 & 0.89 & 0.88 \\
{GSM-$psd$}$_{\rm res}$ & 4 & 0.48 & 0.64 & 0.84 & 0.85 \\
\end{tabular}
\label{tab:SF-mirror-comparison}
\end{ruledtabular}
\end{table}

The results shown in Table~\ref{tab:SF-mirror-comparison} indicate 
that the continuum couplings significantly reduce the  SF for the knockout of well-bound nucleons from dripline nuclei. 
In order to understand the underlying mechanism,
the squared spectroscopic amplitude  of the  $0p_{3/2}$ resonant state are listed in the last row of Table~\ref{tab:SF-mirror-comparison} and
the squared spectroscopic amplitudes ${{\cal A}^2_{p_{3/2}}(k_p)}$ of the non-resonant states are shown in the supplemental material \cite{SM}. Since the orbital
$0p_{3/2}$ is well bound, the $p_{3/2}$ continuum is not expected to contribute. Indeed the value of the SF is determined by the contribution from the $0p_{3/2}$  bound pole, which is, however, significantly reduced compared to the HO-SM prediction, see Table~\ref{tab:SF-mirror-comparison}.

\begin{figure}[!htb]
\includegraphics[width=\linewidth]{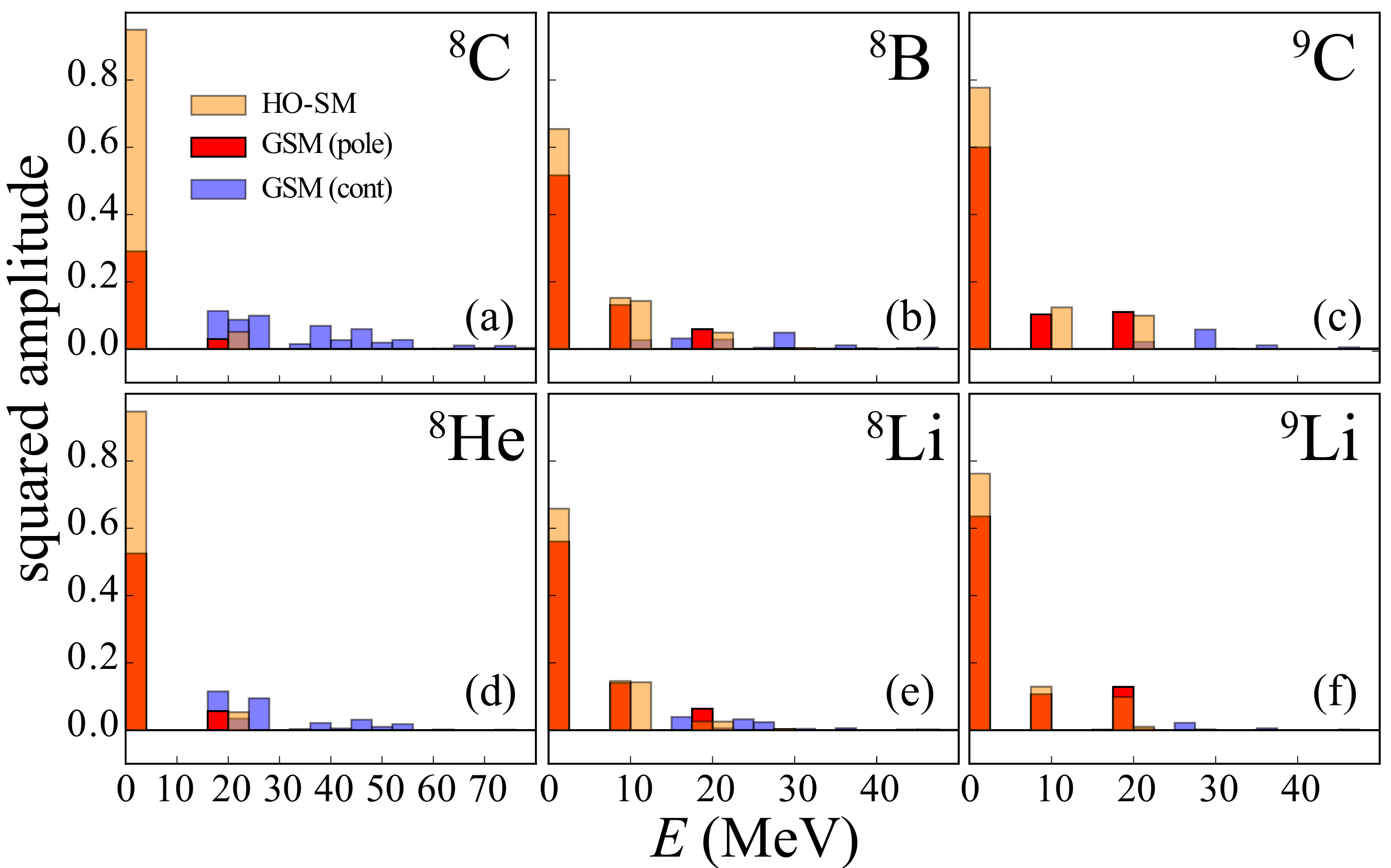}
\caption{Squared HO-SM and GSM amplitudes of  shell-model configurations for  $^{8}$C (a), $^8$B (b), $^{9}$C (c), $^{8}$He (d), $^8$Li (e), and   $^9$Li (f). GSM calculations were performed in $psd$ space with $N_{\rm cont}$=4. The contributions from the GSM pole space and scattering continuum space is shown separately.
The configuration energy is defined as the sum of  s.p. energies of valence nucleons relative to the SM configuration with the lowest energy. Note that the GSM amplitudes are generally complex, hence only their real parts are shown.}
\label{fig:all_config_distributions}
\end{figure}
To understand this reduction, Fig.\,\ref{fig:all_config_distributions} compares the HO-SM  wave-function decomposition  to that of GSM-$psd$ ($N_{\rm cont}$=4)
for the six nuclei considered.
It is seen that the continuum plays different roles in $^8$C and $^9$C. For the unbound $^8$C, there is a broad distribution of GSM configurations involving non-resonant states, which are absent in the HO-SM calculations.
For the particle-bound $^9$C, the impact of the continuum on the dominant HO-SM configurations is not as dramatic.

The GSM weights of dominant configurations in the proton-unbound $^8$C are: 29\% for $\pi(0p_{3/2})^4$,  30\% for $\pi(0p_{3/2})^3(p_{3/2}^{\rm cont})$,  
and 21\% for $\pi(0p_{3/2})^2(p_{3/2}^{\rm cont})^2$. For $^9$C, the leading configurations are: $\pi(0p_{3/2})^4\nu(0p_{3/2})$ (60\%),
$\pi(0p_{3/2})^3(p_{1/2}^{\rm cont})\nu(0p_{3/2})$ (10\%),
and $\pi(0p_{3/2})^2(0p_{1/2})^2\nu(0p_{1/2})$ (9\%). It is seen, therefore, that the  proton structure of $^8$C differs significantly from the proton structure of $^9$C. This is related to the reduction 
of the $\nu0p_{3/2}$  bound pole contribution for the former and the quenching of the SF seen in Table~\ref{tab:SF-mirror-comparison}. A similar situation is predicted for the $^9$Li$\rightarrow^8$He+$p$ process. As seen in Fig.\,\ref{fig:all_config_distributions}, the structures of the mirror nuclei $^9$Li and $^9$C are very similar. Since $^8$He is neutron bound,  it has a larger pole contribution than  
$^8$C.

\begin{figure}[!htb]
\includegraphics[width=0.8\linewidth]{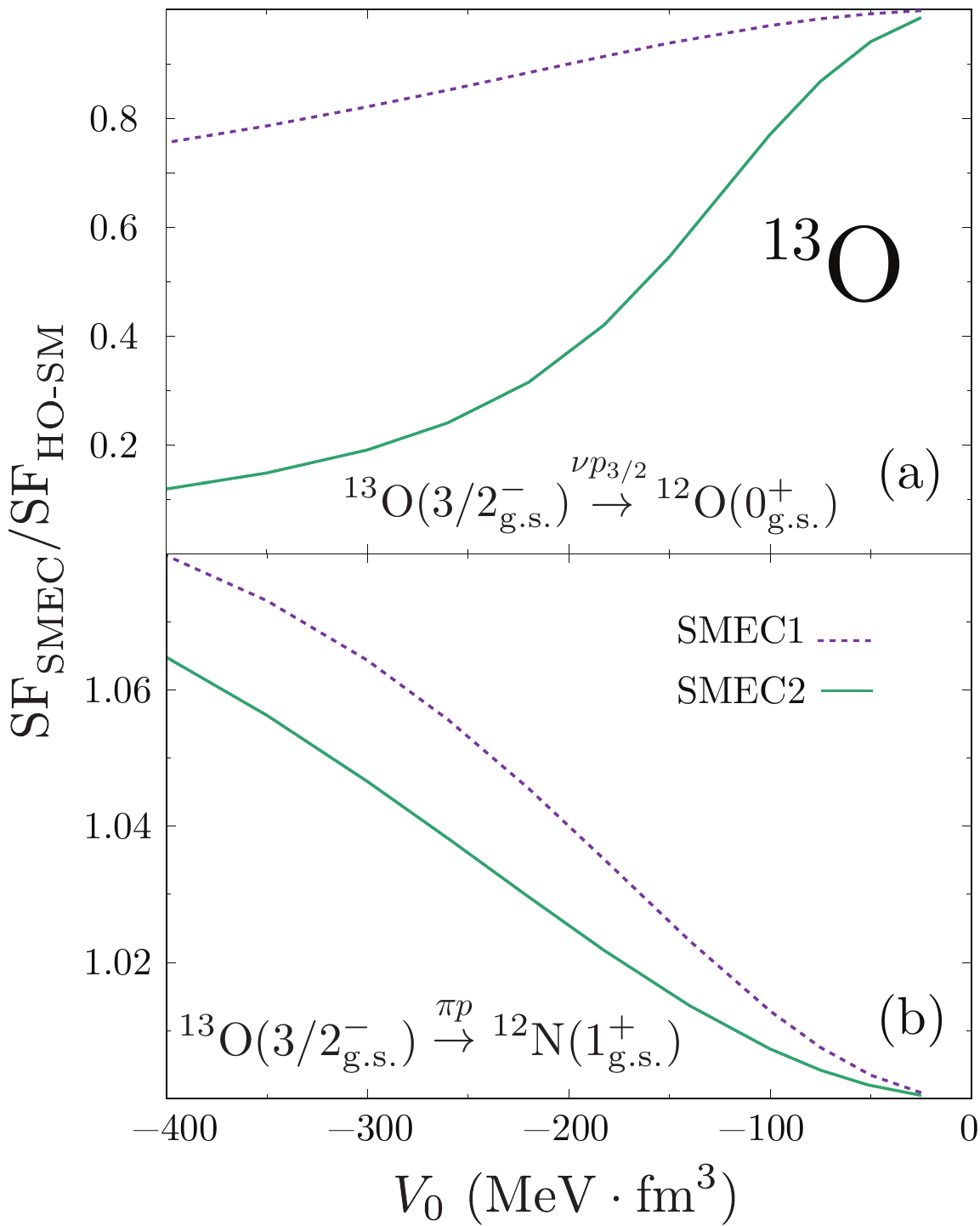}
\caption{The ratio of spectroscopic factors obtained in  SMEC (SMEC1 and SMEC2 variants) and HO-SM for (a) neutron and (b) proton removal from the g.s. of $^{13}$O as a function of the continuum-coupling strength $V_0$. 
}
\label{fig:SMEC_np_SF_13O_V0}
\end{figure}

To further illustrate the quenching of neutron spectroscopic factors in proton-dripline nuclei, in Fig.\,\ref{fig:SMEC_np_SF_13O_V0}(a) we show SMEC results for the one-neutron knockout from the g.s. of $^{13}$O to the unbound g.s. of $^{12}$O, which can  decay by the emission of two protons \cite{Charity2020}. 
The calculations have been carried out by assuming the resonant character of  the ground state of $^{12}$O (SMEC2) and by ignoring the unbound nature of this nucleus (SMEC1).  
Consequently, curve named SMEC2 is calculated by coupling the three lowest $J^{\pi} = 3/2^-$ SM states in $^{13}$O to the channels 
$\left[^{12}{\rm O}(0^+_1)\otimes{\nu}{ p}_{3/2}\right]_{3/2^-}$,  
$\left[^{12}{\rm N}(1^+_1)\otimes{\pi}{ p}_{3/2}\right]_{3/2^-}$, and $\left[^{12}{\rm N}(1^+_1)\otimes{\pi}{ p}_{1/2}\right]_{3/2^-}$. Furthermore, the four lowest $0^+$ SM states in $^{12}$O are coupled to the channels $\left[^{11}{\rm N}(1/2^+_1)\otimes{\pi}{ s}_{1/2}\right]_{0^+}$ and $\left[^{11}{\rm O}(3/2_1^-)\otimes{\nu} p_{3/2}\right]_{0^+}$.
In the curve SMEC1, the resonance character of the ground state of $^{12}$O has been neglected. One can see that opening of the proton emission channel in SMEC2 leads to a dramatic decrease of  the neutron SF.  This is consistent  with the GSM results for $^9$C.

{\it Removal of weakly-bound or unbound  nucleons from dripline systems}.
We begin from the GSM analysis of 
 SFs for the removal of a majority species nucleon from $^9$C and $^9$Li. 
As seen in  Table~\ref{tab:SF-mirror-comparison},
contrary to the removal of minority species nucleons, the SFs for ${^9{\rm C}}(3/2^-_{\rm g.s.}) \rightarrow \pi p_{3/2} + {^8{\rm B}}(2^+_{\rm g.s.}) $ and $ {^9{\rm Li}}(3/2^-_{\rm g.s.}) \rightarrow \nu p_{3/2} + {^8{\rm Li}}(2^+_{\rm g.s.})$  are weakly impacted by  the continuum coupling, in spite of the fact that the contribution to the SFs from the scattering $p_{3/2}$ space increases, see Table~\ref{tab:SF-mirror-comparison} and the supplemental material \cite{SM}. 
This behavior is due to the small separation energy, as the wave functions of  valence nucleons have a broad spatial distribution. As a result, the mother nucleus can be viewed in terms of a weak coupling of the valence nucleon to a daughter nucleus core,
which means that the nucleon-removal  process has little impact on the core \cite{Tostevin2021,Hebborn2021}. This spectator approximation is nicely seen in the wave function amplitudes of mother and daughter nuclei in Fig.~\ref{fig:all_config_distributions}.

The large SFs for the removal of weakly bound or unbound  nucleons are also seen in  the SMEC calculation. Figure\,\ref{fig:SMEC_np_SF_13O_V0}(b) illustrates the SF of  $ {^{13}{\rm O}}(3/2_{\rm g.s.}^-) \rightarrow {^{12}{\rm N}}(1^+_{g.s.}) $ $\ell=1$ proton decay. 
Here, the removal of a proton yields the weakly bound ($S_p = 0.6$\,MeV) g.s. of $^{12}$N. 
The curve SMEC2 is obtained by coupling  the three lowest $J^{\pi} = 3/2^-$ SM states in $^{13}$O to the channels $\left[^{12}{\rm N}(1^+_1)\otimes{\pi}{ p}_{3/2}\right]_{3/2^-}$, $\left[^{12}{\rm N}(1^+_1)\otimes{\pi}{ p}_{1/2}\right]_{3/2^-}$, and
$\left[^{12}{\rm O}(0^+_1)\otimes{\nu}{ p}_{3/2}\right]_{3/2^-}$.
 The four lowest $1^+$ states in $^{12}$N are coupled to the channels $\left[^{11}{\rm N}(1/2^+_1)\otimes{\nu}{ s}_{1/2}\right]_{1^+}$, $\left[^{11}{\rm C}(3/2^-_1)\otimes{\pi}{ p}_{1/2}\right]_{1^+}$, and $\left[^{11}{\rm C}(3/2_1^-)\otimes{\pi} p_{3/2}\right]_{1^+}$. As in Fig. 2a, the resonance character of $^{12}$N has been neglected in curve SMEC1.
Interestingly,  the SF slightly increases with increasing continuum-coupling strength. This is opposite to what was found for the neutron SFs (see Fig.\,\ref{fig:SMEC_np_SF_13O_V0}(a)) but is consistent with the GSM results.
The small difference between SMEC1 and SMEC2 results signifies that including the coupling to the closed channels has no significant effect on  one-nucleon SFs.

\begin{figure}[!htb]
\includegraphics[width=0.8\linewidth]{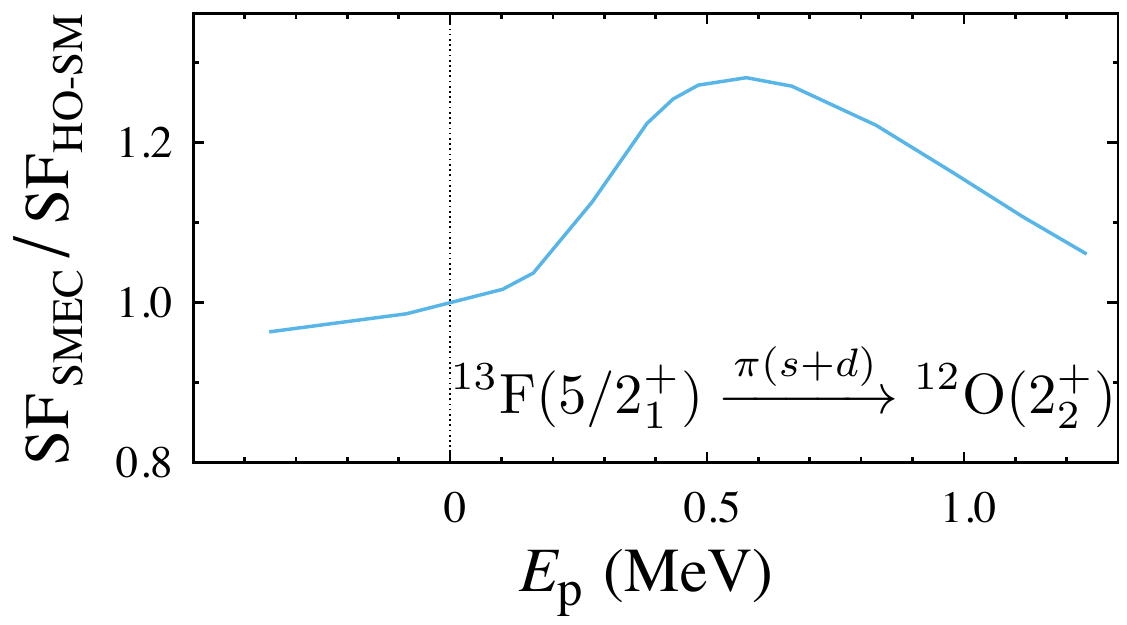}
\caption{Ratio between SFs obtained in SMEC2 and HO-SM for the  $5/2^+_1$ proton resonance in  $^{13}$F as a function of one-proton decay energy. Two partial proton waves, $s_{1/2}$ and $d_{5/2}$, primarily contribute to the SF. The dotted line at $E_{p} = 0$\,MeV denotes the proton decay threshold [$^{12}$O$(2_2^+) \otimes \pi$].}
\label{fig:SF_13F}
\end{figure}

A proton decay of a resonance in $^{13}$F  with a width $\Gamma = 1.01(27)$\,MeV constitutes another excellent example of a large SF associated with a removal of an unbound nucleon from a dripline system.  The exotic $^{13}$F nucleus, recently observed in Ref.~\cite{Charity2021}, is located four neutrons beyond the proton drip line. The resonance in question, placed 0.48(19)\,MeV above the proton decay threshold of $^{12}$O$(2_2^+)$, was tentatively identified as the $5/2_1^+$ excited state. 
Due to the unbound character  of $^{13}$F, the SF of the $5/2_1^+$ resonance has been calculated using SMEC2. 
In this calculation, all open channels are included, namely: $\left[ ^{12}{\rm O}(2^+_2)\otimes{\pi} s_{1/2} \right]_{5/2^+}$,
$\left[ ^{12}{\rm O}(2^+_2)\otimes{\pi} d_{5/2} \right]_{5/2^+}$, $\left[ ^{12}{\rm O}(0^+_1)\otimes{\pi} d_{5/2}\right]_{5/2^+}$, $\left[ ^{12}{\rm O}(2^+_1)\otimes{\pi} s_{1/2} \right]_{5/2^+}$, $\left[ ^{12}{\rm O}(2^+_1)\otimes{\pi} d_{5/2} \right]_{5/2^+}$, and $\left[ ^{12}{\rm O}(0^+_2)\otimes{\pi} d_{5/2}\right]_{5/2^+}$.
The continuum coupling strength $V_0 = -100$\,MeV$\cdot$fm$^3$ reproduces the measured decay width. As one can see in Fig.\,\ref{fig:SF_13F},  the  ratio of SFs calculated in  SMEC and HO-SM is large; its maximum appears close to the suggested experimental energy of the $5/2_1^+$ resonance with respect to the one-proton decay threshold $[^{12}$O$(2_2^+) \otimes \pi]$ \cite{Charity2021}. As we discussed above, this enhanced SF means that the wave function of the daughter nucleus is  weakly coupled to the valence proton. This is not surprising as the decaying  resonance lies very  close to the threshold; hence, its wave function is threshold-aligned due to the continuum coupling  \cite{Okolowicz2020}.


{\it Summary}.
Using two different formulations of the shell model for open quantum systems, we have demonstrated and explained a  non-intuitive result that the  continuum-coupling effect on SFs  is large for the removal process of a well-bound nucleon but is weak when the removed particle is weakly bound or unbound. 
This behavior can be naturally explained within the continuum SM  in terms of coupling to the non-resonant space.
{\it When a minority species nucleon is removed}, the daughter nucleus moves in the direction of the dripline. This leads to
an appreciable change in  configurations of weakly-bound nucleons that are impacted by continuum effects; thus, the SF is reduced.
For instance, in the cases considered, the daughter nuclei $^8$C and $^{12}$O are proton-unbound, and $^8$He is a 4$n$ halo. 
{\it When a majority species nucleon is removed}, the daughter nucleus moves away from the dripline and 
stays closer to the core of the parent system. Consequently, based on the spectator approximation, one expects the SF to be  large.

The continuum couplings for nucleons in weakly-bound orbits depend  on the number of particle continua (in GSM), the resonance nature of states involved, or  (in SMEC) on the number of decay channels included and the value of the continuum coupling strength. Moreover, as shown in SMEC, the difference between the spectroscopic factor calculated in SMEC and in the SM depends on the isospin structure of the  interaction:
 for large $|S_n - S_p|$ the asymmetry appears between the nucleon-nucleon interaction in weakly-bound and well-bound systems and  the interaction between unlike nucleons becomes reduced \cite{GSMbook}.  
 The effect depends on the angular momentum involved. For large $\ell>3$, the asymmetry in a removal of minority or majority  nucleon is expected to be reduced. Future theoretical studies should answer which of these two ingredients (continuum coupling or interaction effects)  prevail in different mass regions of the nuclear chart.


{\it Acknowledgements}.---
Discussions with Robert Charity, Alexandra Gade, and Lee Sobotka are gratefully acknowledged.
This material is based upon work supported by
the U.S. Department of Energy, Office of Science, Office of
Nuclear Physics under Awards No.DE-SC0013365 (Michigan
State University),  No.DE-SC0018083 (NUCLEI SciDAC-4 collaboration), 
by the National Natural Science Foundation of China, the Chinese Academy of Sciences and Peking University under grants
No.NPT2020KFY13 (the State Key Laboratory of Nuclear Physics and Technology, Peking University), No.11975282 (the National Natural Science Foundation of China), No. XDB34000000 (the Strategic Priority Research Program of Chinese Academy of Sciences), No.XDPB15 (the Key Research Program of the Chinese Academy of Sciences) and by the COPIN and COPIGAL French-Polish scientific exchange programs.

\end{document}